\documentclass[final,5p,times,twocolumn]{elsarticle}

\usepackage{hyperref}

\usepackage{upgreek}

\journal{Journal of \LaTeX\ Templates}

\begin{document}

\begin{frontmatter}

\title{Description and performance results of the trigger logic of TUS and Mini-EUSO to search for Ultra-High Energy Cosmic Rays from space}

\author[label1,label2]{M.~Bertaina\corref{cor}}
\cortext[cor]{Corresponding author}
\ead{bertaina@to.infn.it}
\author[label1,label2,label3]{D.~Barghini}
\author[label2]{M.~Battisti}
\author[label4,label5]{A.~Belov}
\author[label1,label6]{M.~Bianciotto}
\author[label1,label2]{F.~Bisconti}
\author[label7]{C.~Blaksley}
\author[label6]{K.~Bolmgren}
\author[label8,label9]{G.~Cambi\`e}
\author[label10]{F.~Capel}
\author[label7,label8,label9]{M.~Casolino}
\author[label7]{T.~Ebisuzaki}
\author[label1,label2,label3]{F.~Fenu}
\author[label12]{M.A.~Franceschi}
\author[label6]{C.~Fuglesang}
\author[label1,label2]{A.~Golzio}
\author[label13]{P.~Gorodetzky}
\author[label14]{F.~Kajino}
\author[label5]{P.~Klimov}
\author[label1,label2]{M.~Manfrin}
\author[label8,label9]{L.~Marcelli}
\author[label15]{W.~Marsza\l}
\author[label2]{M.~Mignone}
\author[label1,label2]{H.~Miyamoto}
\author[label12]{T.~Napolitano}
\author[label13]{E.~Parizot}
\author[label8,label9]{P.~Picozza}
\author[label17]{L.W.~Piotrowski}
\author[label15]{Z.~Plebaniak}
\author[label13]{G.~Pr\'ev\^ot}
\author[label8,label9]{E.~Reali}
\author[label12]{M.~Ricci}
\author[label7]{N.~Sakaki}
\author[label5]{S.~Sharakin}
\author[label15]{J.~Szabelski}
\author[label7]{Y.~Takizawa}
\author[label15]{M.~Vrabel}
\author[label5]{I.~Yashin}
\author[label5]{M.~Zotov}

\address[label1]{Dipartimento di Fisica, Universit\`a di Torino, Torino, Italy}
\address[label2]{INFN, Sezione di Torino - Torino, Italy}
\address[label3]{INAF, Osservatorio Astrofisico di Torino - Torino, Italy}
\address[label4]{Faculty of Physics, M.V. Lomonosov Moscow State University - Moscow, Russia}
\address[label5]{Skobeltsyn Institute of Nuclear Physics, Lomonosov Moscow State Univ. - Moscow, Russia}
\address[label6]{KTH Royal Institute of Technology - Stockholm, Sweden}
\address[label7]{RIKEN - Wako, Japan}
\address[label8]{INFN, Sezione di Roma Tor Vergata - Roma, Italy}
\address[label9]{Dipartimento di Fisica, Universit\`a di Roma Tor Vergata, Roma, Italy}
\address[label10]{Technical University of Munich - Munich, Germany}
\address[label12]{INFN, Laboratori Nazionali di Frascati - Frascati, Italy}
\address[label13]{APC, Univ Paris Diderot, CNRS/IN2P3, CEA/Irfu, Obs de Paris, Sorbonne Paris Cit\'e, France}
\address[label14]{Konan University, Kobe, Japan}
\address[label15]{National Centre for Nuclear Research - Lodz, Poland}
\address[label17]{Faculty of Physics, University of Warsaw - Warsaw, Poland}

\vspace{-1.cm}

\begin{abstract}
The Trigger Logic (TL) of the Tracking Ultraviolet Setup (TUS) and
Multiwavelength Imaging New Instrument for the Extreme Universe Space Observatory (Mini-EUSO) space-based projects of the Joint Experiment 
Missions - EUSO (JEM-EUSO) program is summarized. The performance results on the search for Ultra-High Energy Cosmic Rays (UHECRs) are presented.

\end{abstract}


\end{frontmatter}



TUS~\cite{BARGHINI2021} and Mini-EUSO~\cite{Bacholle_2021} are 
aimed at demostrating the detection principle of UHECRs from space in view of
the planned K-EUSO~\cite{universe8020088} and POEMMA~\cite{Olinto_2021}
missions. TUS operated in 2016--2017 as a part of the Lomonosov 
satellite orbiting at 500~km from ground while Mini-EUSO is operational since 2019 on the ISS. Both telescopes are 
based on an Fresnel optical system (mirrors for TUS and lenses for Mini-EUSO) which focus near-UV light (290 -- 430 nm) on an array of 
PhotoMultiplier Tubes (256 PMT channels for TUS and 2304 pixel channels for Mini-EUSO). Both instruments adopt a multi-mode TL with 
sampling time (ST) ranging from $\upmu$s to tens of ms 
to search for UHECRs and slower phenomena occurring in the atmosphere such as 
Transient Luminous Events (TLEs), meteors and dark matter~\cite{refid0}. These 
TLs are fine-tuned versions of the one designed for a large space-based  
mission~\cite{ABDELLAOUI2017150} 
similarly to what is done for the balloon missions EUSO-SPB1/2~\cite{BATTISTI2019349,FILIPPATOS2021}.
\begin{figure*}[!ht]
\includegraphics[width=1.\linewidth]{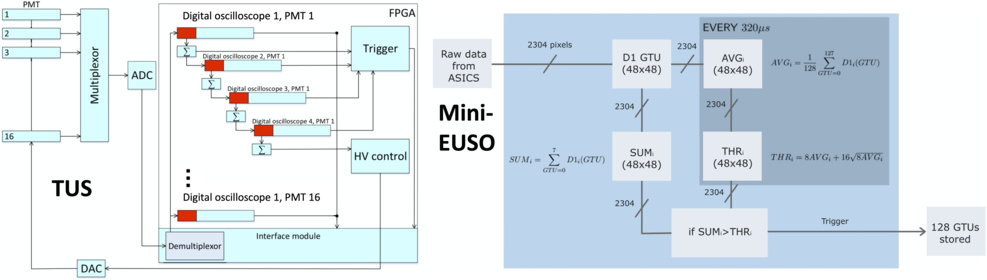}
\caption{Block schemes of TUS (left) Photo Detector Module (PDM) electronics (cluster of 16 PMTs) and of 
Mini-EUSO (right) 1st level TL.}
\label{fig:tus_minieuso_logics}
\end{figure*}

A description of the TUS acquisition logic is reported in~\cite{Klimov2017} and 
Fig.~\ref{fig:tus_minieuso_logics}. The readout operates in 4 modes specifically designed 
for different classes of events. However, they can not operate simultaneously, therefore, the operation mode 
is defined at start run. Normally, it is changed every few weeks or months.
The Extensive Air Shower (EAS) mode is aimed at the detection of UHECRs. The ST is 0.8 $\mathrm 
\upmu$s. 
The other modes are designed for TLEs and meteor detection and have a ST of 25.6 $\upmu$s (or 0.4 ms) 
and 6.6 ms, respectively. A software TL is structured in 2 steps to allow background rejection and the acceptance of EAS events. 
A fast ADC digitizes the signals in each time frame (TF). The digitized signals are summed up on a sliding 
window of 16 TFs for each PMT. The integrated values are then compared with a preset threshold (this
is done at PDM FPGA level). 
If 
the threshold is overcome, the 1st level TL is activated. 
The 2nd level TL is a pixel mapping TL implemented in the central processor board, 
which acts as a
contiguity trigger. This procedure selects cases of sequential triggering of spatially contiguous
pixels that are also adjacent in time, allowing for the selection of events with
different spatial-temporal patterns. The preset value of the number of neighbouring pixels (N) 
sequentially activated by a signal from a given event is set to $N$ = 3 but can be changed by command 
during flight.
Once the persistency is longer than $N$, the 2nd level TL is issued, the 
acquisition procedure is stopped and data transfer is started. For 50-60 s the detector is in dead time 
before restarting the acquisition. 
A description of the Mini-EUSO TL is reported in~\cite{BELOV20182966} and
Fig.~\ref{fig:tus_minieuso_logics}. Mini-EUSO adopts the photon counting technique. The data 
acquisition operates simultaneously at 3 different STs and stores the data in 3 different timescales (D1, D2 and D3).
The 2.5 $\upmu$s ST (D1 data) is the fastest one and has a dedicated TL based on signal excess at pixel level of at least 
16$\sigma$ above the average background level due to nightglow or other natural and/or artificial sources on ground and/or in atmosphere. The 
excess is estimated integrating 8 consecutive TFs of D1
while the average background level is updated every 320 $\upmu$s. The system can acquire up to 4 packets of 128 D1 TFs every 5.24 s, then it is in dead time till the end of the 5.24 s period. 
Each pixel operates independently and the 8 D1 integration time matches the time required by a light signal to cross diagonally the 
pixel's field of view 
at ground (like 16 TFs in TUS).
This is the TL designed for fast signals like
UHECRs or fast TLEs. The 2nd TL operates at 320~$\upmu$s ST (D2) and has a dedicated 
TL similar to the 1st TL 
which runs in parallel. Up to 4 packets of 128 D2 data can be acquired every 5.24 s. D2 TL is suitable for TLEs.
Finally a continuous data taking at 41 ms
ST (D3) is performed. Data are grouped in blocks of 128 D3 TFs corresponding to 5.24 s. These data are used offline to produce
UV maps needed to compute the exposure for UHECR observation, and search for slow events (i.e. meteors). Prior to flight 
the TL was successfully tested in laboratory and in open-sky conditions~\cite{Bisconti2022}. 

The expected performance of TUS and Mini-EUSO for EAS detection was tested by means of 
ESAF simulations~\cite{BARGHINI2021,BELOV20182966}. 
The sensitivity of both detectors turned out to be around (for TUS) or above (for Mini-EUSO) 
10$^{21}$ eV.
TUS collected $\sim$7$\times$10$^4$ and Mini-EUSO so far $\sim$5$\times$10$^4$ events in EAS mode. Among
them a few hundreds had characteristic light curve with a pronounced maximum and full duration at 
half-maximum from 40 to 80 $\upmu$s, which is quite consistent with the simulated detector response to the EAS signal. However,
the amplitude of all EAS-like events corresponds to UHECR energies well above 10$^{21}$ eV.
Moreover, 
the majority of EAS-like events were registered above continents, several times close by airports. 
Mini-EUSO revealed their association with ground flashers due to their repetitive occurrence. 
The non repetitive ones
were excluded from an EAS origin by comparing at the same time their light profile and track image with simulated EAS. 
TUS detected a few events characterized by a moving light spot. The most interesting one was registered above 
US~\cite{Khrenov_2020}, however, an UHECR origin
is highly unlikely. TUS collected a total geometric exposure of 
 $\sim$1550 km$^2$ sr yr in EAS mode, while the current estimation for Mini-EUSO is of $\sim$400 km$^2$ sr yr.
 The amount of events of different nature collected by TUS and Mini-EUSO
demonstrate a multifunctionality of an orbital fluorescent observatory and its
usefulness for various astrophysical and geophysical studies, and provide an invaluable experience 
in view
of K-EUSO and POEMMA. 

\section*{Acknowledgments}
This work was supported by ROSCOSMOS, by ASI 
agreements ASI-INAF n.2017-14-H.O and ASI-INFN n.2020-26-Hh.0, by 
CNES, by NCN (Poland) grant 2017/27/B/ST9/02162, 
by the ISE School of Moscow University
``Fundamental and Applied Space Research''
and it is based on research
materials 
carried out in the space experiment ``UV atmosphere''.

\bibliography{TUS-Mini-EUSO-logics}

\end{document}